\documentclass[twocolumn]{aastex631}
\usepackage{multirow}
\usepackage{xspace}
\usepackage{graphicx}
\usepackage{booktabs}
\usepackage[T1]{fontenc}

\usepackage{amsmath}

\newcommand{\Msun}{\,{\rm M_\odot}}
\newcommand{\fedd}{\,{f_{\rm Edd}}}
\newcommand{\alphaox}{\,{\alpha_{\rm ox}}}
\newcommand{\Mblack}{M_\bullet}

\newcommand{\Ha}{\rm H\alpha}
\newcommand{\koral}{\texttt{KORAL}\xspace}
\newcommand{\heroic}{\texttt{HEROIC}\xspace}

\begin{document}

\title{Mildly Super-Eddington Accretion Onto Slowly-Spinning Black Holes \\ Explains the X-Ray Weakness of the Little Red Dots}

\author[0000-0001-9879-7780]{Fabio Pacucci}
\affiliation{Center for Astrophysics $\vert$ Harvard \& Smithsonian, 60 Garden St, Cambridge, MA 02138, USA}
\affiliation{Black Hole Initiative, Harvard University, 20 Garden St, Cambridge, MA 02138, USA}

\author[0000-0002-1919-2730]{Ramesh Narayan}
\affiliation{Center for Astrophysics $\vert$ Harvard \& Smithsonian, 60 Garden St, Cambridge, MA 02138, USA}
\affiliation{Black Hole Initiative, Harvard University, 20 Garden St, Cambridge, MA 02138, USA}



\begin{abstract}
JWST has revealed a population of low-luminosity AGN at $z>4$ in compact, red hosts (the ``Little Red Dots'', or LRDs), which are largely undetected in X-rays. We investigate this phenomenon using GRRMHD simulations of super-Eddington accretion onto a SMBH with $\Mblack=10^7\Msun$ at $z\sim6$, representing the median population; the SEDs that we obtain are intrinsically X-ray weak. The highest levels of X-ray weakness occur in SMBHs accreting at mildly super-Eddington rates ($1.4<\fedd<4$) with zero spin, viewed at angles $>30^\circ$ from the pole. X-ray bolometric corrections in the observed $2-10$ keV band reach $\sim10^4$ at $z=6$, $\sim5$ times higher than the highest constraint from X-ray stacking. Most SEDs are extraordinarily steep and soft in the X-rays (median photon index $\Gamma=3.1$, mode of $\Gamma=4.4$). SEDs strong in the X-rays have harder spectra with a high-energy bump when viewed near the hot ($>10^8$ K) and highly-relativistic jet, whereas X-ray weak SEDs lack this feature. Viewing a SMBH within $10^\circ$ of its pole, where beaming enhances the X-ray emission, has a $\sim1.5\%$ probability, matching the LRD X-ray detection rate. Next-generation observatories like AXIS will detect X-ray weak LRDs at $z\sim6$ from any viewing angle. Although many SMBHs in the LRDs are already estimated to accrete at super-Eddington rates, our model explains $50\%$ of their population by requiring that their masses are overestimated by a mere factor of $\sim3$. In summary, we suggest that LRDs host slowly spinning SMBHs accreting at mildly super-Eddington rates, with large covering factors and broad emission lines enhanced by strong winds, providing a self-consistent explanation for their X-ray weakness and complementing other models.
\end{abstract}


\section{Introduction} 
\label{sec:intro}

X-ray photons have been foundational to studying compact objects. For example, the first X-ray source detected outside the Solar System (Sco X-1, \citealt{Giacconi_1962}, a highly accreting neutron star, \citealt{Shklovsky_1967}), the first confirmed stellar-mass black hole (Cygnus X-1, \citealt{Bowyer_1965}) and the first confirmed supermassive black hole (SMBH, 3C 273, \citealt{Schmidt_1963}) are all powerful X-ray sources. 
Searching for black holes in the high-energy sky is a fruitful method because X-ray sources are sparse ($\sim 500 \, \rm deg^{-2}$, in the Chandra Source Catalog, \citealt{CSC_2024}). Hence, an X-ray detection of a centrally located source within a galaxy has been the ``gold standard'' to assess the presence of a SMBH.

A newly discovered, enigmatic population of galaxies in the high-$z$ Universe is challenging this fact: the ``Little Red Dots'' (LRDs). LRDs are a population of high-$z$ sources detected in many JWST deep fields \citep{Labbe_2023, Kocevski_2023, Harikane_2023, Maiolino_2023_new, Matthee_2023}, with the bulk of their distribution in the redshift range $4 < z < 8$, and a median of $z \sim 6$  \citep{Kocevski_2024, Akins_2024, Kokorev_2024_census}.

A few properties of the LRDs are apparent. They have very red near-infrared colors \citep{Matthee_2023} and are very compact, with a median effective radius of $r_e \sim 150 \, \rm pc$ \citep{Baggen_2023, Labbe_2023}, while some are even smaller with $r_e < 35 \, \rm pc$ \citep{Furtak_2023_lensed}. This property would lead to extreme stellar densities in their galactic cores \citep{Pacucci_2024_z_evolution}. Additionally, these sources are abundant compared to the previously known population of high-$z$ AGN. With a number density of $10^{-4} -10^{-5} \, \rm Mpc^{-3} \, mag^{-1}$ at $z \sim 5$, they are $10-100$ times more numerous than the faint end of quasar luminosity function \citep{Greene_2023}. Hence, if they contain central SMBHs, they belong to a previously unexplored and fainter population of AGN.

Given the properties of the LRDs, their physical interpretation bifurcates into two paths. The LRDs are either very massive, star-forming, and compact galaxies (see, e.g., \citealt{Labbe_2023}), or they contain a central SMBH, typically in the mass range $\Mblack \sim 10^6-10^8 \Msun$ (see, e.g., \citealt{Harikane_2023, Maiolino_2023_new}). 

A majority of these sources do show telltale signs of the presence of an active SMBH. \cite{Greene_2023} found that $\sim 60\%$ of the LRDs investigated display definitive evidence of broad emission lines, with $\Ha$ FWHMs of $> 2000 \, \rm km \, s^{-1}$: a signature of Type-1 AGN, for which an estimate of the mass can be derived \citep{Greene_2005}. The absence of broad forbidden lines (e.g., the [OIII]5007) suggests that such emission cannot be associated with outflows \citep{Maiolino_2024_Xray}. If the LRDs contain a central SMBH, their SEDs are assembled by a dust-obscured AGN and a young, blue galaxy \citep{Kocevski_2023}. \cite{PG_2024} argue that many LRDs are extremely intense and compact starburst galaxies, which produce a significant amount of dust; their global energy output is dominated by emission from OB stars and an obscured AGN.

Surprisingly, most LRDs observed thus far are undetected in the X-rays, leading to the ``X-ray weakness problem'', which starkly contrasts with other AGN-related properties of the LRDs.
For example, in a comprehensive sample of $341$ LRDs spanning several JWST fields, \cite{Kocevski_2024} found only two X-ray detected LRDs, which are moderately obscured with $\log_{10}\,(N_{\rm H}/{\rm cm}^{2}$) of $22.72$ and $23.3$.
Although the LRDs account for only $\sim 10\%$ of the AGN population discovered by JWST \citep{Maiolino_2024_Xray}, their lack of X-ray emission is compelling, as it adds to their already puzzling properties.

Previous studies systematically searched for X-ray emissions in LRDs. \cite{Yue_2024_Xray} used archival Chandra data on 19 spectroscopically selected LRDs with broad emission lines and detected none. The X-ray stacking analysis also led to non-detection, with an inferred soft-band ($0.5-2$ keV) upper limit that is $\sim 1$ dex lower than what is expected from typical Type-1 AGN. They also suggest that extremely large absorbing column densities, i.e., $\log_{10}\,(N_{\rm H}/{\rm cm}^{2}) > 24$, would be required to entirely suppress the X-ray emission, especially since Chandra is probing energies of $\sim 2.5 - 50$ keV at a representative redshift of $z = 5$. Although the X-ray obscuration does increase with redshift \citep{Peca_2023}, it is unlikely that very broad $\Ha$ lines are associated with such substantial absorbing column densities. 
As pointed out by \cite{Yue_2024_Xray}, some broad absorption line (BAL) AGN show values of $\log_{10}\,(N_{\rm H}/{\rm cm}^{2}) > 24$ \citep{Blustin_2008}; however, the spectra of LRDs do not show typical BAL features \citep{Greene_2023}. Additionally, Type 1 AGN with $\log_{10}\,(N_{\rm H}/{\rm cm}^{2}) > 24$ are very rare in the local Universe, i.e., $\lesssim 5\%$ of the whole population \citep{Ananna_2022}.
Considering these constraints, \cite{Yue_2024_Xray} suggest that LRDs may be intrinsically X-ray weak, as also pointed out by \cite{Inayoshi_2024}.

\cite{Maiolino_2024_Xray} also executed an X-ray stacking analysis of a large JWST sample of Type-1 and Type-2 AGN, leading to widespread non-detections. Their stacking analysis led to upper limits of $1-2$ dex lower than expected from standard AGN emission. Contrarily to \cite{Yue_2024_Xray}, they suggest that dust-poor clouds from the broad-line region (BLR) with sufficiently large column densities and abnormally large covering factors can explain the X-ray weakness. They also indicate that steep X-ray spectra, such as those observed in Narrow Line Seyfert 1 (NLSy1) galaxies, can contribute to explaining an intrinsic X-ray weakness \citep{Vasudevan_2007}.

\cite{Ananna_2024} performed a stacking analysis of a sample of $21$ LRDs detected behind the lensing galaxy cluster Abell 2744. This study also led to widespread non-detections and an upper limit on the X-ray emission, with constraints similar to those reported by \cite{Yue_2024_Xray}.

To conclude, \cite{King_2024} suggested that LRDs may be highly-beamed sources, fed at or above the Eddington rate, similar to the case of ultraluminous X-ray sources (ULXs). The resulting gas outflow velocities of $\sim 0.1-0.2c$ would make virial mass estimates difficult and lead to an overestimate. Interestingly, the Galactic X-ray binary Cyg X-3, which may be viewed as a ULX from observers located along the axis of the funnel, shows signs of X-ray weakness \citep{Cyg_X3_2023}. Additionally, based on semi-empirical models of the emission in different accretion regimes, \cite{Lupi_2024} also suggested that the masses of the SMBHs in the LRDs may be overestimated due to accretion at largely super-Eddington rates.

In this Letter, we demonstrate that the LRDs' X-ray weakness can be explained by SMBHs accreting at mildly super-Eddington rates, with SEDs significantly different from those of radiatively efficient, ``standard'' AGN. Intrinsically soft X-ray spectra, redshifted out of Chandra's range, can elegantly explain the lack of high-energy detections while supporting the other hallmarks of the black hole hypothesis (e.g., broad lines and significant covering factors).
For the first time, we support this interpretation with a suite of GRRMHD (General Relativistic Radiation Magnetohydrodynamics) simulations of SMBHs accreting above the Eddington rate, with various values of the spin parameter and viewing angles.
In Sec. \ref{sec:primer}, we provide a primer on different accretion regimes for black holes. In Sec. \ref{sec:numerical_methods}, we describe our numerical codes for the GRRMHD simulations and ray-tracing. 
Then, we present our results in Sec. \ref{sec:results}. Finally, in Sec. \ref{sec:conclusions}, we summarize our findings and conclude.

\section{A Primer on Accretion Regimes}
\label{sec:primer}

The Eddington ratio $\fedd$ is defined as the ratio between the mass accretion rate $\dot{\Mblack}$ onto the black hole and the corresponding Eddington rate:
\begin{equation}
    \fedd = \frac{\dot{\Mblack}}{\dot{M}_{\rm Edd}} \, ,
\end{equation}
where $\dot{M}_{\rm Edd}$ is the mass accretion rate at which a relativistic thin accretion disk \citep{Novikov_Thorne} would radiate at the Eddington luminosity. 
$\fedd$ is the main parameter that defines the accretion regime and, ultimately, the characteristics of the accretion disk, jet, and the shape of the emerging spectral luminosity.

For $10^{-2} \lesssim \fedd \lesssim 1$, a geometrically-thin, optically thick, radiatively efficient accretion disk is present \citep{SS_1973,Novikov_Thorne}\footnote{The inner parts of the disk are geometrically thick and somewhat radiatively inefficient already by $\fedd\approx0.5$, but for simplicity we take the limit to be $\fedd\approx1$.}. This accretion mode is often referred to as the ``standard disk'' model and is the most widely studied for black holes spanning the entire mass range. The ``bias'' towards this accretion regime is likely caused by the combination of two effects: black holes accreting in this regime are sufficiently bright to be detected but, at the same time, do not require extremely large (thus, infrequent at $z=0$) reservoirs of cold, accretable gas. In the radiatively efficient thin disk regime, typical matter-to-energy conversion factors $\epsilon$ are of the order $\epsilon \sim 0.1$ (depending on the spin, \citealt{Bardeen_1970,Novikov_Thorne}), where the emitted luminosity is $L = \epsilon \dot{M}c^2$.

The standard thin disk regime is itself sub-divided into two sub-regimes. For $\fedd\lesssim0.1$, the disk is gas-pressure dominated and is thermally stable. However, for $0.1\lesssim\fedd\lesssim1$, radiation pressure dominates, and the disk becomes viscously and thermally unstable \citep{Lightman_1974,Shakura_Sunyaev_1976,Jiang_2013}. Unfortunately, the unstable regime overlaps with the classic luminosity range of quasars, which prevents self-consistent modeling of these astrophysically important systems.

For $\fedd \lesssim 10^{-2}$ and $\fedd \gtrsim 1$, the disk becomes radiatively inefficient. In the lower range ($\fedd \lesssim 10^{-2}$), an extremely hot, two-temperature, optically-thin, advection-dominated accretion flow (ADAF) is formed, which is geometrically thick and has typical values of $\epsilon \ll 0.1$ \citep{Narayan_1994, Narayan_1995, Abramowicz_1995, Narayan_2008, Yuan_Narayan_2014}. What little radiation is present in this accretion state has hardly any effect on the dynamics of the optically thin gas. Sgr A* is a classic example of a SMBH accreting in ADAF mode  \citep{Yuan_2003}. Low $\fedd$ systems often accumulate a strong magnetic field at small radii around the black hole to form a Magnetically Arrested Disk \citep[MAD,][]{Igumenshchev_2003, Narayan_2003_MAD}. In this state, they produce relativistic jets extending to thousands of gravitational radii $r_g = 2G\Mblack/c^2$ if the black hole is spinning \citep{Tchekhovskoy_2011}. It is likely that most black holes, at least in the local Universe, accrete in the ADAF mode and are in the MAD state \citep{Narayan_2022}.
The accretion mode alternative to the MAD state is aptly named SANE (Standard And Normal Evolution).

For $\fedd \gtrsim 1$, the accretion disk is radiatively inefficient because of photon trapping \citep{Begelman_1979,Abramowicz_1988}. The disk is geometrically thick, prompting the name ``slim disk'' for these systems. Crucially, the slim disk regime is thermally stable \citep{Abramowicz_1995,Narayan_1995,Chen_1995}, which facilitates theoretical studies, especially using numerical simulations \citep{Sadowski_2014,McKinney_2014}. Jets are prominent if the disks reach the MAD state and if the black holes are spinning \citep{Narayan_2017,Curd_2023,Ricarte_2023}. Such systems will be extremely X-ray-bright for pole-on observers because of relativistically beamed radiation from the hot disk and jet \citep{Curd_2019}.

Radiation plays a critical role in both the dynamics and thermodynamics of standard and slim disks (i.e., for the entire range $\fedd \gtrsim 0.01$). There is a strong coupling between the gas and the radiation, while radiative pressure inflates the disk in the vertical direction as $\fedd$ approaches or exceeds unity \citep{Abramowicz_1988}. Additionally, jets and winds driven only by radiation can occur in super-Eddington systems \citep{Sadowski_2014}.
For these reasons, a GRRMHD code, i.e., a GRMHD code with a self-consistent treatment of the radiation field, is required. The work reported in this paper uses the GRRMHD code \koral \citep{Sadowski_2013,Sadowski_2014}.

This primer has highlighted the physical differences characterizing the three fundamental accretion regimes. GRRMHD-derived spectral models are available in the ADAF regime ($\fedd\lesssim0.01$) and, albeit more scarcely, in the super-Eddington regime ($\fedd\gtrsim1$). However, accretion rates in between are challenging to simulate. For $\fedd\lesssim0.1$, disks are geometrically very thin and require extremely high resolution, making them challenging for GRRMHD codes. On the other hand, for $\fedd\gtrsim0.1$ and especially as $\fedd\to1$, although disks are geometrically thicker and potentially accessible to GRRMHD codes, they are thermally unstable, as already mentioned \citep{Shakura_Sunyaev_1976}. Only a handful of full GRRMHD simulations have been successfully run in this accretion regime; in these few simulations, strong magnetic fields stabilize the accretion flow \citep{Sadowski_2016} and a curious ``puffy disk'' structure is observed (\citealt{Lancova_2019,Wielgus_2022}). The models so far are limited to non-spinning black holes.

In the present work, we focus on the super-Eddington regime ($\fedd\gtrsim1$), where the accretion disk is stable, and we have both GRRMHD codes to simulate the gas magnetohydrodynamics and radiation field of the disk \citep[e.g.,][]{Sadowski_2013,Sadowski_2014} as well as post-processing codes to compute spectra \citep[e.g.,][]{Hero_2015,Heroic_2016}. The results we report here are, however, valid only for the super-Eddington regime and cannot be extended to  thermally unstable $\fedd\lesssim1$ systems. The latter regime, though of great interest for modeling traditional quasars, is beyond the reach of our codes. Simulating them would require entirely different simulation techniques (that do not yet exist) and theoretical assumptions.

\section{Numerical Methods}
\label{sec:numerical_methods}
Calculating the spectral luminosity emitted by SMBHs accreting in the super-Eddington regime involves a two-stage process. First, we conducted GRRMHD simulations of the accretion flow for specific model parameters; then, we solved for the radiation field using a post-processing code. The two steps are described in their generalities in the following subsections; the interested reader is referred to the original studies describing the GRRMHD code \koral \citep{Sadowski_2013, Sadowski_2014}, the radiation post-processing code \heroic \citep{Hero_2015, Heroic_2016}, as well as their application to the case of ultra-luminous X-ray sources \citep{Narayan_2017} and tidal disruption events \citep{Curd_2019,Curd_2023}.

\subsection{GRRMHD Simulations with \koral}

The super-Eddington accretion process onto SMBHs is simulated using the GRRMHD code \koral \citep{Sadowski_2013, Sadowski_2014}, which incorporates the effects of gas dynamics, magnetic fields, and radiation within a fixed gravitational field.
These simulations are performed with the Kerr metric in Kerr-Schild coordinates and employ the M1 closure method \citep{Levermore_1984} to handle radiative transfer, along with a radiative viscosity term to address limitations inherent in the M1 scheme \citep{Sadowski_2015}.
The simulations include radiative processes such as synchrotron emission, free-free and bound-free \citep[taken from][]{Sutherland_Dopita_1993} emission and absorption, and Compton scattering. 
All these simulations are performed in 3D to correctly resolve the magnetorotational instability \citep{Balbus_1991}.

Simulations begin with weakly magnetized gas in an equilibrium torus orbiting the SMBH; the parameters are adjusted to achieve the desired mass accretion rate \citep{Narayan_2022}, which, in our case, varies between mildly to strongly super-Eddington (see Sec. \ref{subsec:sims}). The topology of the initial magnetic field is appropriate for MAD models \citep{Tchekhovskoy_2011}. MAD-type accretion disk models around SMBHs are characterized by strong magnetic fields, significantly affecting the accretion process. These disks occur when the magnetic field becomes so intense that it disrupts the inflow of gas onto the SMBH, creating a region where magnetic forces dominate the dynamics of the disk. 

Our simulations are run on a grid with $256\times192\times32$ cells in $(r,\theta,\phi)$, where $\phi$ extends over only a $\pi/2$ wedge (hence the effective resolution is 128 cells over $2\pi$). Each simulation is run up to a time of $30000 t_g\approx 410$ hrs, where the gravitational time scale for a black hole of mass $\Mblack$ is $t_g = G\Mblack/c^3 \approx 50$ s for $\Mblack = 10^7 \Msun$. By the end of the simulation, the disk is in a steady state out to roughly $\approx 50 r_g$ (see \citealt{Narayan_2017,Ricarte_2023} for additional details). Once the system has reached a steady state, the spectra are calculated via ray tracing from the last $5000 t_g$.

The standard procedure of a $\sigma=1$ cut was applied to the GRRMHD runs. This technique addresses problematic regions in GRRMHD simulations where the quantity $\sigma = B^2/(4\pi \rho c^2)$, the ratio of magnetic stress to rest mass energy density, exceeds unity. In these regions, which commonly coincide with the highly magnetized funnel, the density becomes very low, causing issues as numerical errors accumulate \citep{Gammie_2003}; hence, cutting out these areas makes the simulation reliable and more accurate. This method is commonly used in various GRMHD simulations, including those used for interpreting the Event Horizon Telescope observations (see, e.g., \citealt{EHT_2019,Tsunetoe_2021}).

\subsection{Radiation Post-Processing with \heroic}

Post-processing of the radiation field is carried out using the general relativistic \heroic code \citep{Hero_2015, Heroic_2016}, which solves the detailed radiation field based on data from the GRRMHD simulations. \heroic accounts for the angular distribution of radiation in the local fluid frame and spans a wide range of frequencies, from radio to gamma rays.
Since its development, \heroic has been enhanced to better handle bremsstrahlung in the relativistic regime, utilizing methods developed from \cite{Narayan_1995} and spectral distribution from \cite{Gould_1980}. Additionally, it uses opacity tables from the \texttt{CHIANTI} database \citep{Dere_1997, Landi_2013, Del_Zanna_2015} as well as from \citet{Sutherland_Dopita_1993}, for temperatures below $10^8$ K, and includes a relativistic Comptonization module for high temperatures \citep{Jones_1968, Coppi_1990}. \heroic also incorporates thermal synchrotron emission and absorption \citep{Narayan_1995, Mahadevan_1996} and can handle two-temperature plasmas \citep{Sadowski_2017}.

In the present work, post-processing involves two stages. First, \heroic takes the time-averaged (over the final $5000t_g$) and axisymmetrized data from \koral for the gas density, temperature, and four-velocity, as well as the magnetic field strength. To be conservative, it caps the gas temperature\footnote{This temperature cap prevents numerical errors in isolated cells in \koral from causing artificially high temperatures that affect the inverse Compton scattering in the output SEDs. The calculated spectra hardly change relative to not capping the temperature, but we do it for safety.} at $10^9$ K, which is the typical coronal temperature in Seyfert galaxies (see, e.g., \citealt{Akylas_2021}). Using this information and the appropriate Kerr spacetime metric, the code iteratively solves for the frequency-dependent radiation field at each grid point, including all the emission, absorption, and scattering opacities and the effect of relativistic Comptonization. At the conclusion of this stage, the code generates a source function at each position on the grid. Using this source function, in the second stage \heroic performs general relativistic ray-tracing, again with all opacities and scattering included, to compute the observed spectra from different viewing angles. Note that \heroic only calculates the continuum spectrum. It does not model relativistically broadened X-ray iron lines or related reflection phenomena \citep[e.g.,][]{Reynolds_2023}.

The observed SEDs are re-normalized so that their bolometric luminosity matches the value expected from the Eddington ratio of each run, given a fixed black hole mass of $10^7 \, \Msun$. This task required a consistent re-normalization across all runs of a factor of $10$ for the radiation field, suggesting that \koral overestimated the temperature field by a factor of $10^{1/4} \approx 1.8$. This procedure is standard practice to correct discrepancies in calculating the radiation and temperature fields performed by \koral, using the M1 closure method. A similar radiation and temperature field matching was also applied to other studies based on \koral and \heroic, e.g., \cite{Narayan_2017}. We also note that the fact that a consistent re-normalization was required for all runs suggests that our simulations are consistent with each other.

\subsection{Description of the Simulations}
\label{subsec:sims}
Our study involves 12 GRRMHD simulations, all assuming a SMBH mass of $10^7\Msun$. This mass is representative of the sample of SMBHs discovered thus far in LRDs at $z > 4$ \citep{Harikane_2023, Maiolino_2023_new}. For example, the sample of LRDs recently analyzed in \cite{Maiolino_2024_Xray} is characterized by a median mass of $\Mblack = 3 \times 10^7 \Msun$.

The GRRMHD simulations vary the Eddington ratio, ranging from mildly super-Eddington (i.e., $\fedd = 1.4$) to strongly super-Eddington (i.e., $\fedd = 13.4$).
Three values of the dimensionless SMBH spin parameter $a$ are considered: 0, 0.68, 0.9. By definition, $a$ is required to lie in the range $0 \leq |a| \leq 1$, where $a = 0$ is a
Schwarzschild (i.e., non-rotating) black hole, and $a = 1$
is a maximally rotating black hole (although the limiting value $a=1$
is not achievable in practice, \citealt{Thorne_1974}).

These 12 simulations are then post-processed to obtain the emerging spectral luminosity for 8 different viewing angles $i$ of the observer relative to the pole. The inclination angle varies between $i=10^\circ$ to $i=80^\circ$, in intervals of $10^\circ$. Hence, a grid of $12$ simulations $\times$ $8$ angles leads to a total of $96$ SEDs, which are used to investigate the physical parameter spaces relevant to our problem. A summary of the physical properties of the 12 simulations performed is presented in Table \ref{tab:sims}.

\begin{table}[h]
    \centering
    \caption{Properties of the 12 GRRMHD simulations performed. The SED emerging from each simulated system is analyzed from 8 different viewing angles: $10^\circ$ to $80^\circ$ from the pole.}
    \begin{tabular}{cccc}
        \toprule
        \# & Black Hole Mass [$M_{\odot}$] & Spin Parameter a & $\fedd$ \\
        \midrule
        1  & $10^7$ & 0.9  & 13.4 \\
        2  & $10^7$ & 0.9  & 6.0  \\
        3  & $10^7$ & 0.9  & 2.8  \\
        4  & $10^7$ & 0.9  & 2.4  \\
        5  & $10^7$ & 0.9  & 1.4  \\
        6  & $10^7$ & 0.68 & 9.3  \\
        7  & $10^7$ & 0.68 & 4.2  \\
        8  & $10^7$ & 0.68 & 2.6  \\
        9  & $10^7$ & 0.68 & 1.4  \\
        10 & $10^7$ & 0.0  & 8.4  \\
        11 & $10^7$ & 0.0  & 3.5  \\
        12 & $10^7$ & 0.0  & 2.4  \\
        \bottomrule
    \end{tabular}
    \label{tab:sims}
\end{table}

\section{Results}
\label{sec:results}
In this Section, we present the results of our study. We first summarize the X-ray properties of our super-Eddington SEDs and show how their \textit{observed} X-ray emission depends on four parameters: Eddington ratio, spin, inclination, and redshift. Then, we apply these findings to explain the X-ray weakness of the population of SMBHs detected in the LRDs and how a slight overestimate of a factor of $\sim 3$ in their mass measurement could lead the majority of them to the mildly super-Eddington regime investigated here.
Super-Eddington accretion leads naturally to X-ray weakness, lower spin, large covering factors, and broad emission lines due to strong winds.

\subsection{X-ray Bolometric Corrections}
\label{subsec:K_X}
We begin by investigating what fraction of the bolometric luminosity of a given SED is emitted in the $2-10$ keV X-ray band. Hence, we calculate the X-ray bolometric correction in the rest frame of the SMBH:
\begin{equation}
    k_X = \frac{L_{\rm bol}}{L_{\rm 2-10 \, keV}^{z=0}} \, .
\end{equation}
Values of $k_X$ for standard AGN are in the range $\sim~10-50$, except for very bright quasars (i.e., with $L_{\rm bol}>10^{47} \, \rm erg \, s^{-1}$) for which it can reach values of $\sim 100$ \citep{Duras_2020}. Note that a larger value of $k_X$ indicates a weaker X-ray source.

\begin{figure*}%
    \centering
\includegraphics[angle=0,width=0.49\textwidth]{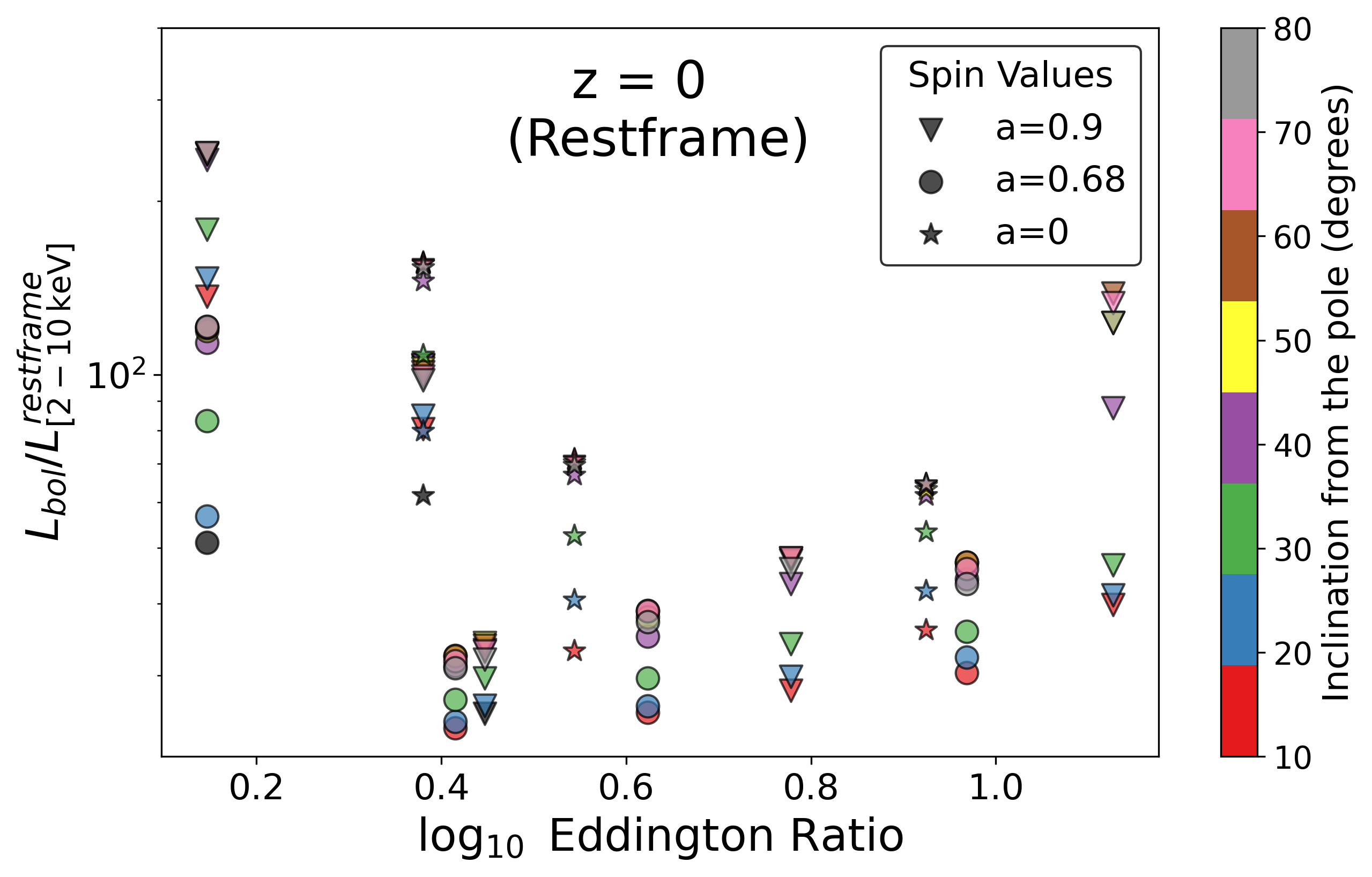} \hfill \includegraphics[angle=0,width=0.49\textwidth]{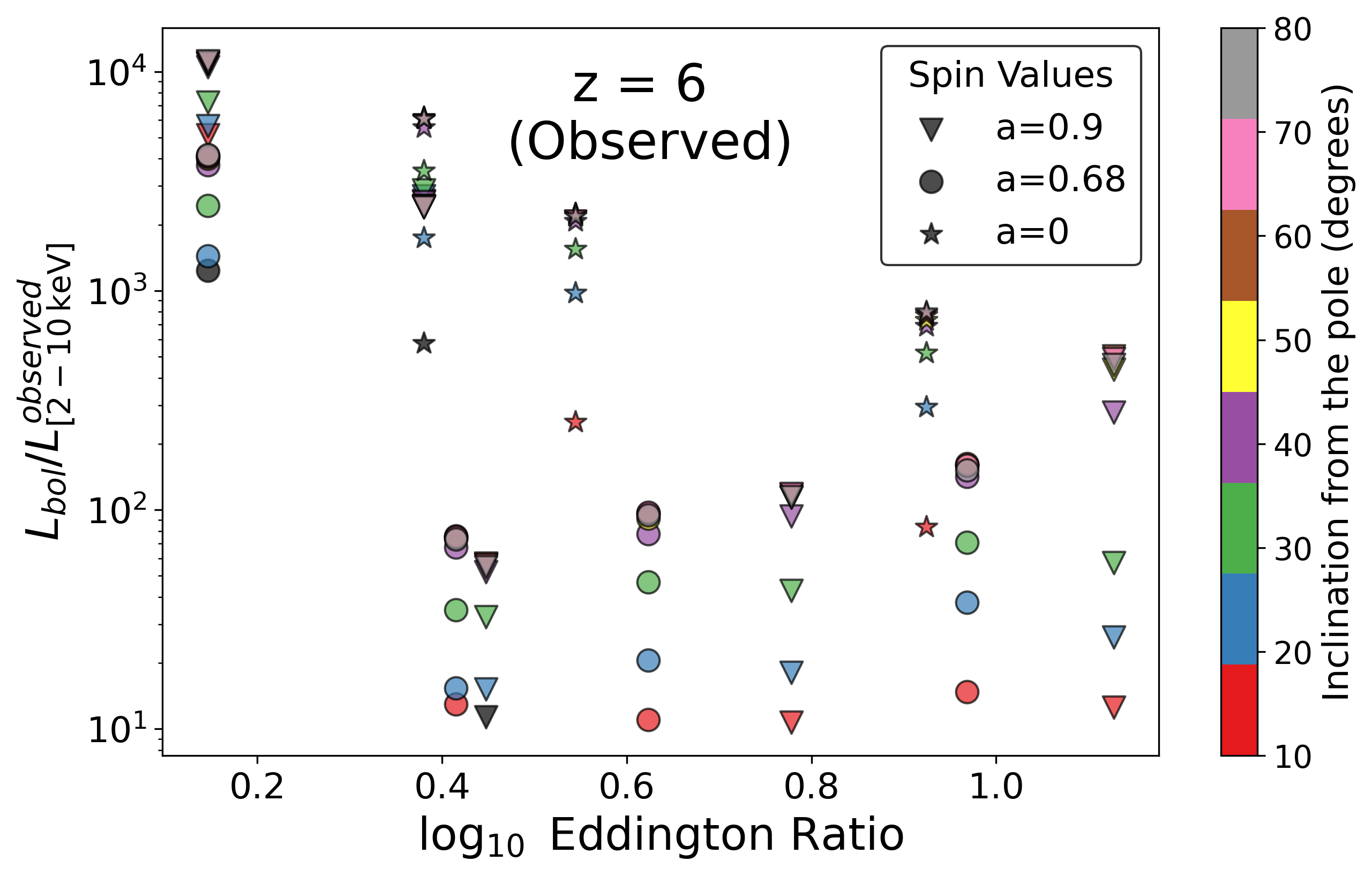}
    \caption{Comparison of the $2-10$ keV bolometric correction as a function of the logarithm of the Eddington ratio in the restframe (left panel) and in the \textit{observed} $2-10$ keV band when the source placed at the median redshift of $z=6$ (right panel). Different colors indicate varying inclination angles from the pole (degrees), and different markers represent different spin values: triangles for $a=0.9$, circles for $a=0.68$, and stars for $a=0$. Mildly super-Eddington accretion rates ($1.4<\fedd<4$) onto slowly spinning black holes, if observed at large inclination angles from the pole, lead to extremely large values of the bolometric correction, especially if the SMBH is highly redshifted.}
    \label{fig:kx_comparison}%
\end{figure*}

If the emitted SED is soft in the X-ray, we expect a portion of the spectrum to be redshifted out of the observed $2-10$ keV energy band for $z \gg 0$, corresponding to a restframe band $[2-10](1+z)$ keV. Hence, we are also interested in the \textit{observed} X-ray bolometric correction calculated at $z$, which we define as:
\begin{equation}
    k_X^{z=6} = \frac{L_{\rm bol}}{L_{\rm 2-10 \, keV}^{z=6}} \, .
\end{equation}
We chose a representative redshift of $z = 6$ as the median value computed from the most comprehensive catalogs of LRDs available to date (\citealt{Kocevski_2024}, where the median redshift is $z = 6.1$, \citealt{Akins_2024}, where it is $z = 6.5$, and \citealt{Kokorev_2024_census}, where it is $z = 5.6$). We note that using a value of $z \sim 5$, which is closer to the median redshift of the sources presented in \cite{Harikane_2023} and \cite{Maiolino_2023_new}, does not change our results in any significant way.

Our results for the restframe (left) and $z=6$ (right) bolometric corrections are displayed in Fig. \ref{fig:kx_comparison}, as a function of the Eddington ratio, spin value and inclination angle from the pole.
This analysis suggests the following general properties:
\begin{itemize}
    \item Mildly super Eddington accretion, with $1.4<\fedd<4$, leads to the highest bolometric corrections (i.e., they are the most X-ray weak).
    \item SEDs generated by zero spin SMBHs have higher bolometric corrections than those with larger spins\footnote{For zero spin, we ran a simulation with a target $\fedd = 1.4$, but instead of settling down to a steady state, the accretion rate secularly declined below Eddington, and the disk became unstable. This model is not included in Table~\ref{tab:sims}.}. 
    \item Large inclinations from the pole lead to higher bolometric corrections. On the contrary, observing the accretion from a viewing angle close to the pole increases the X-ray flux.
\end{itemize}

As the LRDs are high-$z$ sources, we are primarily interested in the \textit{observed} bolometric corrections calculated for a source at the median redshift of $z=6$; they are displayed in the right panel of Fig. \ref{fig:kx_comparison}.
While the general trends with Eddington ratio, spin, and inclination angle do not change, the values of the bolometric corrections are substantially larger, indicating an astounding \textit{observed} X-ray weakness. 
In particular, $2/3$ of all the SEDs with mildly super-Eddington rates (i.e., $\fedd < 4$) are characterized by $k_X^{z=6} > 10^3$, with some of them even reaching $k_X^{z=6} > 10^4$. These values can explain even GN-28074 \citep{Juodzbalis_2024_extreme}, the most extreme X-ray weak AGN found to date by JWST (see Fig. \ref{fig:LRD_Xray_BC}).

The significant variations in the X-ray bolometric corrections (at a fixed redshift) are due to a combination of different factors. Generally, X-rays are produced by hot gas with temperature $T>10^7$\,K. In the present simulations, such gas is found exclusively in the jet, both at the jet base and along the walls of the funnel. Higher spin values lead to a more prominent jet \citep{Blandford_2019}; correspondingly, non-spinning black holes are the most X-ray-quiet. Jet emission is beamed, hence the observed X-ray emission is strongly dependent on the inclination angle. The Eddington ratio determines the properties of the accretion disk, with higher temperatures being associated with higher values of $\fedd$, which leads to a higher level of X-ray emission. Section \ref{subsec:disk_structures} describes how the structure of the accretion flow determines the observed X-ray emission.

In summary, GRRMHD simulations of super-Eddington accreting SMBHs show that the observed SEDs are soft in the X-rays, especially for mildly super-Eddington accretion ($1.4<\fedd<4$) on slowly spinning black holes when observed at large inclination angles from the pole. If these SEDs, which are already \textit{intrinsically X-ray weak}, are observed at $z \sim 6$, they reach staggering values of the \textit{observed} X-ray bolometric correction, $>2$ orders of magnitude higher than (restframe) reference values for standard AGN.

A note on terminology is warranted. In this work, an "intrinsically X-ray weak" system is defined as one where reduced X-ray emission arises from the intrinsic physical properties of the accretion disk, jet, or corona, or due to the influence of structures close to the black hole (within 100 $r_g$) on the emitted radiation. This definition excludes the effects of obscuration from structures at larger radii, such as a torus. Thus, we consider the system within 100 $r_g$ as a whole, where few high-energy photons are produced, leading to an intrinsically weak X-ray spectrum. Relativistic beaming (see Sec. \ref{subsec:X_weakness}) is not included in this definition; while beaming can enhance the observed X-ray flux from a weak system, it cannot fundamentally alter the intrinsic emission to produce an X-ray strong system.

We conclude by pointing out that the behavior of super-Eddington accretion in the SANE regime is generally similar to that of MAD systems with zero spin, both leading to X-ray-weak outcomes. This is consistent with the results from \cite{Curd_2019}, where both SANE ($a=0$) and MAD ($a=0$) models exhibit X-ray weakness, while only the MAD ($a=0.9$) model is X-ray strong. Therefore, the MAD state is not critical for explaining the X-ray weakness observed in the LRDs, and a SANE accretion state would yield similar results. In the MAD case, a low spin must be invoked to reduce X-ray luminosity, but SANE systems do not require this additional constraint.

\subsection{Optical-UV to X-ray Ratios: the $\alphaox$}
\label{subsec:alpha_ox}
We expand on the study of the X-ray weakness by calculating the values of the $\alphaox$  \citep{Tananbaum_1979}. While the bolometric corrections compare one specific energy range to the entire spectrum, the $\alphaox$ is a point-by-point comparison between the optical-UV and the X-ray emission. 

For our purposes, the $\alphaox$ is a valuable tool to investigate how much our super-Eddington SEDs are X-ray weak (in the soft band) compared to other AGN with the same optical-UV luminosity. 
We use the following definition from \cite{Lusso_2010}:
\begin{equation}
    \alphaox = - \frac{\log_{10} \left( L_{2 \, \text{keV}}/L_{2500 \, \text{\AA}} \right)}{2.605} \, ,
\end{equation}
which is characterized by a mean value of $\alphaox \approx 1.5$ for standard high-luminosity AGN \citep{Lusso_2010}.

\begin{figure}%
    \centering
\includegraphics[angle=0,width=0.45\textwidth]{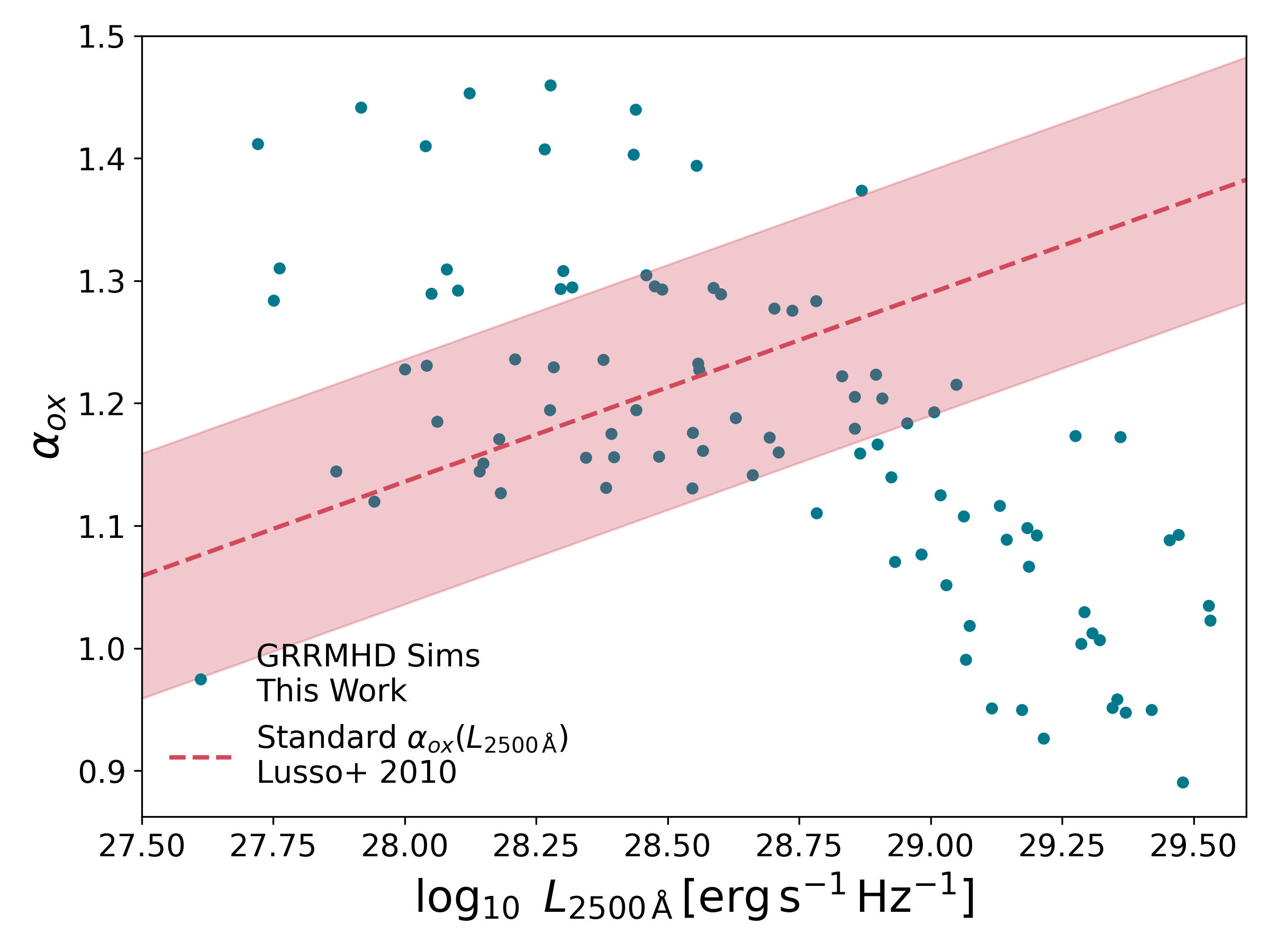} \hfill
\caption{The relationship between $\alphaox$ and $\log_{10} L_{2500}$ for our super-Eddington SEDs. The dashed red line (with its intrinsic scatter) represents the standard $\alphaox(L_{2500 \, \mathrm{\AA}})$ relation from \cite{Lusso_2010}. Many of our super-Eddington SEDs are characterized by values of the $\alphaox$ that are significantly higher, indicating a severe lack of X-ray emission. Our SEDs become X-ray stronger, instead of weaker, with increasing optical-UV luminosity.}
    \label{fig:alpha_ox}%
\end{figure}

Figure \ref{fig:alpha_ox} displays that most of our SEDs are characterized by values of the $\alphaox$ that are significantly different from the standard $\alphaox(L_{2500})$ relation presented in \cite{Lusso_2010}.
In particular, $\sim 55\%$ of our SEDs are outside the dispersion region for typical low-$z$ sources.
In extreme cases, the (restframe) luminosity at $2$ keV is $\sim 20$ times weaker than expected from a standard AGN with the same optical-UV luminosity.
Interestingly, our models show increasing X-ray emission, instead of X-ray weakening, for increasing UV luminosity.

\vspace{0.7cm}
\subsection{Characterizing the X-ray Slope: the Photon Index}
\label{subsec:photon_index}
The photon index, $\Gamma$, characterizes the spectral slope of the power-law component of the SED, and it is a valuable measure to investigate how soft (or hard) an SED is at high energy.
Assuming a power-law component $F_\nu \propto \nu^{-\alpha}$, where $\alpha$ is the spectral index, the photon index $\Gamma$ is usually defined as $\Gamma = \alpha+1$, which leads to:
\begin{equation}
    \nu F_{\nu} \propto \nu^{2-\Gamma} \, .
    \label{eq:gamma}
\end{equation}
Typical values are $1.5<\Gamma<2.5$ \citep{Piconcelli_2005}.

It is crucial to note that $\Gamma<2$ increases $\nu F_{\nu}$ with frequency, while $\Gamma>2$ leads to a decrease. Larger values of $\Gamma>2$ lead to a rapidly steepening SED, which can become extremely soft in the X-rays.

\begin{figure}%
    \centering
\includegraphics[angle=0,width=0.47\textwidth]{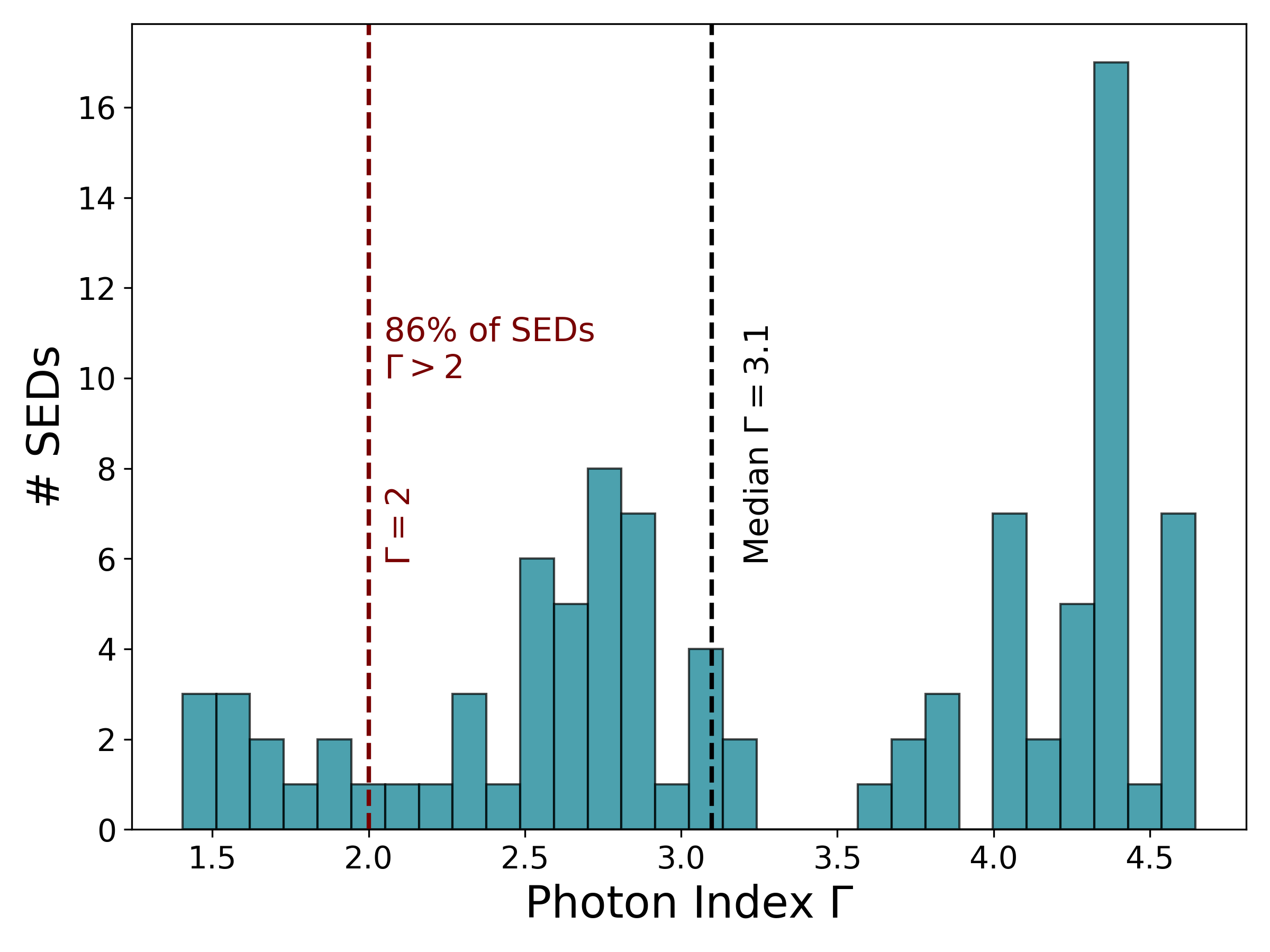} \hfill
    \caption{The photon index distribution for our $96$ super-Eddington SEDs. These values of $\Gamma$ indicate extremely steep, X-ray soft SEDs, declining rapidly with increasing energy.}
    \label{fig:photon_index}
\end{figure}

In Fig. \ref{fig:photon_index}, we display the distribution of photon indexes $\Gamma$ for our 96 SEDs. The photon index is calculated in the observed frame at $2-10$ keV (i.e., $14-70$ keV restframe at $z=6$).
The median value is $\Gamma=3.1$, while the most common is $\Gamma=4.4$. About $86\%$ of our SEDs are characterized by $\Gamma > 2$, which means that $\nu F_{\nu} \propto \nu^{2-\Gamma}$ declines with higher frequency. These values of $\Gamma$ indicate an extremely steep, X-ray soft SED, declining rapidly with increasing frequency. 

Our values of $\Gamma$ resemble the highest among those recently measured in NLSy1 galaxies by eROSITA \citep{Grunwald_2023}.
\cite{Maiolino_2024_Xray} correctly suggested that an intrinsic X-ray weakness in the LRDs could be caused by a steep, extremely X-ray soft SED, similar to that of NLSy1. Our study confirms that this is the case. 
Care needs to be taken when choosing the photon index $\Gamma$ to characterize the (restframe) bolometric corrections of the LRDs. Choosing too low values, especially if $\Gamma < 2$, will lead to overestimating the restframe bolometric correction. Values in the range $\Gamma = 1.7-1.9$, which are typical of sub-Eddington accretion, have been thus far used in the literature to characterize the X-ray weakness of the LRDs \citep{Yue_2024_Xray, Maiolino_2024_Xray, Ananna_2024}.

It is important to note that the spectral descriptors, i.e., $k_X$, $\alphaox$, and $\Gamma$, derived from our GRRMHD simulations are compatible with those recently observed in super-Eddington accreting SMBHs, both in the local and in the high-$z$ Universe.
For example, \cite{Laurenti_2022} studied a sample of low-$z$ AGN ($0.4 \leq z \leq 0.75$) with bolometric luminosities of $\sim10^{46} \rm \, erg \, s^{-1}$ and selected for accretion levels close to Eddington, with $0.9 \leq \fedd \leq 1.2$; they found substantial values of the bolometric correction, ranging from $k_X = 50$ to $k_X = 5100$, and with a median value of $k_X = 175$. They observed, however, smaller values of the photon index, i.e., $\Gamma \sim 2$, in agreement with the values found by another study of local, super-Eddington accreting black holes \citep{Maithil_2024}. However, the values of $\Gamma$ for super-Eddington accreting SMBHs are higher than those for sub-Eddington ones. In the high-$z$ Universe, \cite{Zappacosta_2023} selected highly-accreting AGN at $z > 6$ in the HYPERION sample, finding steep values of the photon index $\Gamma = 2.4 \pm 0.1$, and up to a value of $\Gamma = 3.03^{+1.08}_{-0.89}$, which is compatible with the median value for our SEDs.

\subsection{The Cause of X-ray Weakness: Mildly Super-Eddington, Low Spin, and Beaming}
\label{subsec:X_weakness}
Thus far, we described how super-Eddington SEDs can be extremely X-ray weak globally (with $k_X$), and compared to their optical-UV emission (with $\alphaox$); we have also described how they can be exceptionally soft and steep in the X-rays (with $\Gamma$).
This Section provides a closer look at specific, representative SEDs to detail (i) what parameters cause the X-ray weakness and (ii) how the SEDs are affected by the typical redshift of the LRDs.

\begin{figure*}%
    \centering
\includegraphics[angle=0,width=0.90\textwidth]{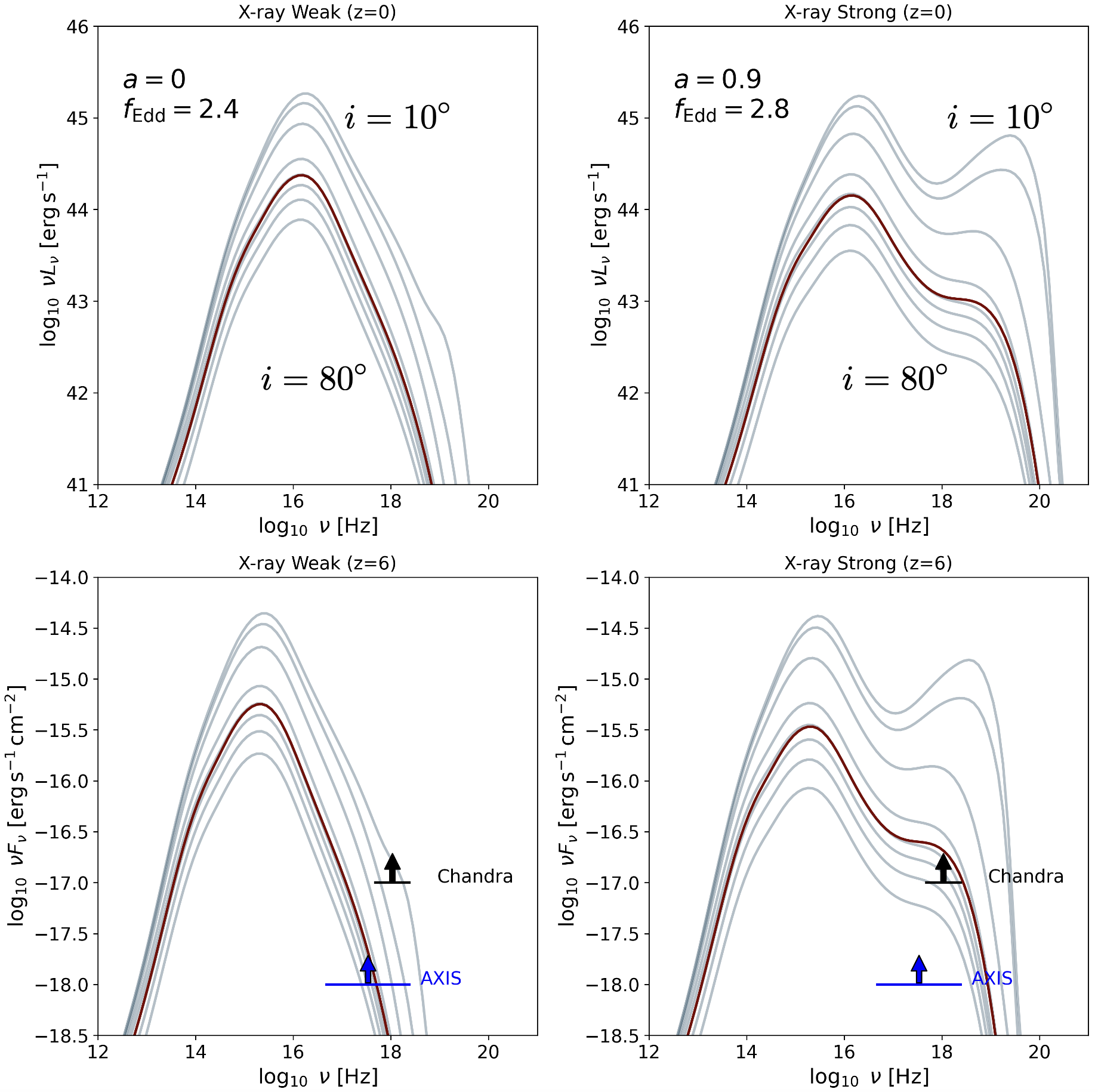} \hfill
    \caption{\textbf{Top Row:} Comparison (at $z=0$) between typical X-ray weak (left, mildly super-Eddington with zero spin) and X-ray strong (right, mildly super-Eddington with high spin) SEDs. \textbf{Bottom Row:} Same as in the top row, but at $z=6$. The 8 gray lines represent the spectrum for lines of sight ranging from $80^\circ$ (bottom) to $10^\circ$ (top) from the pole. The red line indicates the spectrum averaged over $4\pi$ steradians. In the bottom panel, the arrows indicate typical flux limits for deep-field X-ray surveys with Chandra ($2-10$ keV) and AXIS ($0.2-10$ keV). Some of these SEDs extend up to $\sim 1$ MeV.}
    \label{fig:spectra_comparison}
\end{figure*}

Figure \ref{fig:spectra_comparison} compares two representative super-Eddington SEDs, which are named the ``reference SEDs'' for the remainder of this study:
\begin{itemize}
    \item \textbf{X-ray Weak:} SED for a spin $a = 0$ SMBH accreting at a mildly super-Eddington rate ($\fedd = 2.4$).
    \item \textbf{X-ray Strong:} SED for a spin $a = 0.9$ SMBH accreting at a mildly super-Eddington rate ($\fedd = 2.8$).
\end{itemize}
Despite a similar Eddington ratio, the SEDs are markedly different. In particular, the X-ray strong SED is characterized by a much harder X-ray spectrum, with a significant bump extending at (restframe) frequencies of $10^{19} - 10^{20}$ Hz for viewing angles close to the pole (i.e., $<30^\circ$). This high-energy bump is caused by Compton scattering and Lorentz boosting the photons originating from the hottest parts of the accretion flow, as shown in Sec. \ref{subsec:disk_structures}. Even the SED averaged over the solid angle (shown in red) is hard in the X-ray. On the contrary, the X-ray weak SED does not show any high-energy bump, except, minimally, for a viewing angle of $10^\circ$. Hence, the emission is much softer in the X-rays.
The principal difference between the two models is the black hole spin. The $a=0.9$ model has a powerful jet and correspondingly has a large X-ray luminosity. The $a=0$ model has a much weaker jet and is much more X-ray-quiet.

To fully understand how observable these SEDs are with, e.g., the Chandra X-ray observatory, it is necessary to redshift them to a representative redshift for the LRDs, e.g., $z = 6$. This exercise is performed in the bottom panels of Fig. \ref{fig:spectra_comparison}.
A source characterized by an X-ray strong SED can be detectable by Chandra in the observed $2-10$ keV range, assuming a very deep field with a flux limit of $10^{-17} \, \rm erg \, s^{-1} \, cm^{-2}$. In particular, Chandra can detect the mean SED, which is averaged over the solid angle.

The situation is radically different for a source characterized by an X-ray weak SED. Chandra can barely detect only sources observed near the pole, and it is $\sim 2$ orders of magnitude in flux away from detecting the mean SED. At the flux limit of $10^{-17} \, \rm erg \, s^{-1} \, cm^{-2}$, the mean SED is too soft, rendering these sources impossible to detect with Chandra.

The fact that Chandra \textit{may detect} such sources, at the median redshift of $z=6$, only for viewing angles very close to the jet (i.e., $<10^\circ$) leads to a significant corollary. Assuming a SMBH with a collimated jet ejecting from its pole and a random orientation in space, we can calculate the probability that Chandra observes it within an inclination of $10^\circ$ from the pole. The solid angle subtended by a cone with a half-angle of $10^\circ$ is given by $\Omega = 2\pi(1 - \cos \, i)$, where $i$ is the inclination in radians. The probability is then the ratio of this solid angle (multiplied by $2$, to account for the two jets) to the total solid angle of a sphere, i.e., $4\pi$ steradians, resulting in a probability of $\sim 1.5\%$.
Interestingly, \cite{Kocevski_2024} reported the X-ray detection of only $2$ LRDs out of a sample of $341$, leading to a similar detection frequency of $\sim 0.6\%$. 
Thus, it is plausible that Chandra is detecting only the LRDs viewed from minimal angles from the jet of the central SMBH.

Next-generation X-ray observatories, such as AXIS \citep{AXIS_2023, Cappelluti_2023_AXIS}, can detect such weak sources. Assuming an AXIS deep field with a flux limit of $10^{-18} \, \rm erg \, s^{-1} \, cm^{-2}$ \citep{Marchesi_2020, Cappelluti_2023_AXIS}, and an energy range of $0.2-10$ keV, AXIS will detect the mean SED of a typical X-ray weak source at $z\sim 6$ easily and, most likely, from any viewing angle.
Note that the SEDs shown in Fig. \ref{fig:spectra_comparison} belong to sources that are easily detectable by JWST/NIRSpec, assuming a flux limit of $\approx 5.5 \times 10^{-19} \, \rm erg \, s^{-1} cm^{-2}$ for a $3\sigma$ detection at $\sim 1 \, \mu m \approx 3\times 10^{14} \, \rm Hz$ \citep{Pacucci_2023_JWST}.

\begin{figure}%
    \centering
\includegraphics[angle=0,width=0.48\textwidth]{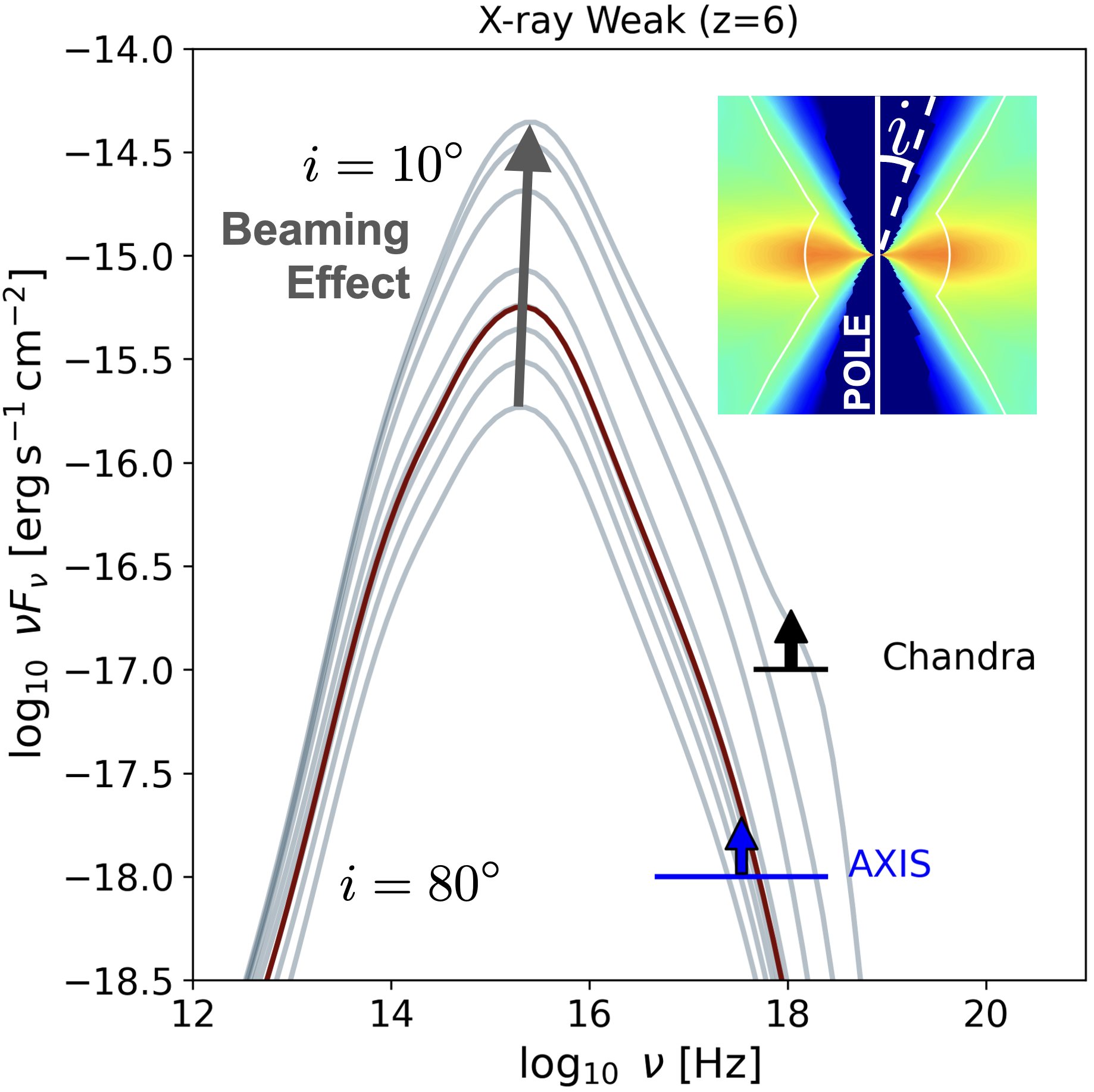} \hfill
    \caption{Effect on the spectrum of a changing line of sight from the pole. The strongest X-ray emission is achieved by looking from a direction close to the pole (i.e., from small inclination angles). The red line indicates the spectrum averaged over $4\pi$.}
    \label{fig:beaming}
\end{figure}

The effect of the viewing angle is essential and warrants a deeper analysis. Figure \ref{fig:beaming} zooms in on the reference X-ray weak SED (at $z=6$) and clarifies the so-called ``beaming effect'' due to observing the emission from lines-of-sight progressively closer to the pole.
In particular, the same SMBH, with the same Eddington ratio and spin, produces SEDs whose peaks differ by $1.5$ dex in flux when the viewing angle changes from $80^\circ$ to $10^\circ$.

Extending the analysis of the SED to lower frequencies, we note that recent studies (e.g., \citealt{Mazzolari_2024}) point out that LRDs that are X-ray weak are also radio-weak.
Regarding the predicted radio luminosity, our simulations do not account for non-thermal jet emission, as both \koral and \heroic assume thermal gas and focus on thermal synchrotron and Bremsstrahlung radiation. Nevertheless, assuming that the radio emission of the jet is proportional to the jet power, $P_{\rm jet}$, we estimate that the X-ray strong model, with $P_{\rm jet} \approx 0.9 \dot{M} c^2$, would be more radio luminous than the X-ray weak model, which has $P_{\rm jet} \approx 0.08 \dot{M} c^2$.

\subsection{X-ray Weak and Strong Accretion Structures}
\label{subsec:disk_structures}

\begin{figure*}%
    \centering
\includegraphics[angle=0,width=0.75\textwidth]{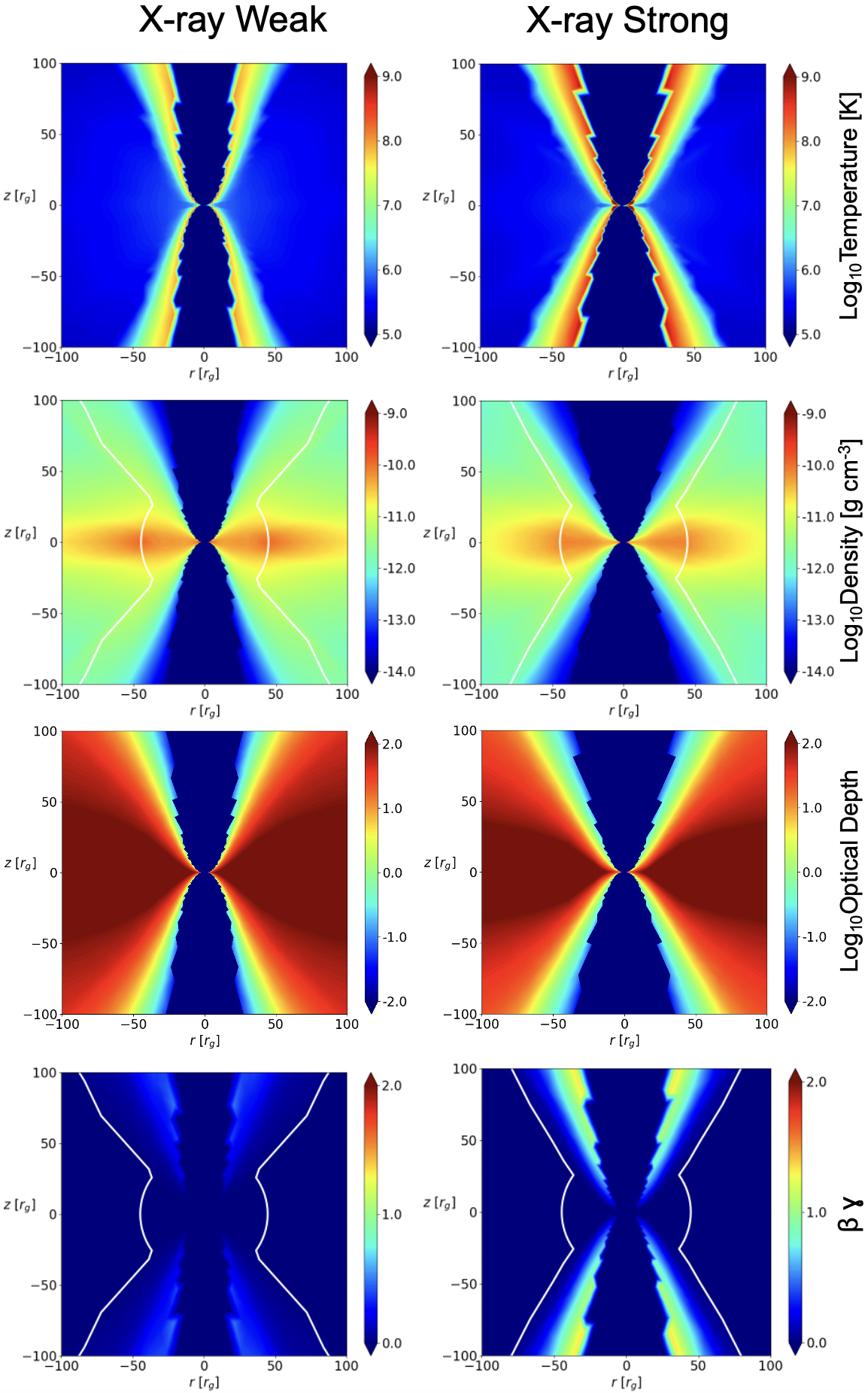} \hfill
    \caption{Comparison between accretion structures in the poloidal plane (the black hole is at $(0,0)$ and the jets are oriented in the vertical direction) for our reference cases of X-ray weak (left) and X-ray strong (right) SEDs (see Sec. \ref{subsec:X_weakness}). The rows are, from top to bottom: temperature, gas density, scattering optical depth, and $\beta \gamma$. X-ray strong accretion structures are significantly hotter and more relativistic.}
    \label{fig:compilation}
\end{figure*}

What are the physical causes of such extreme differences in the X-ray emission? A detailed view of the structure of the accretion flows in the models corresponding to the two reference SEDs (see Sec. \ref{subsec:X_weakness} for their definition) is provided in Fig. \ref{fig:compilation}.
The density distribution in the central panel shows that the accretion disks of X-ray weak and X-ray strong SEDs are similar, with a ``puffed up'' structure typical of super-Eddington accretion.

In general, for sub-Eddington accretion rates, the disk is thinner \citep{Novikov_Thorne, SS_1973}, and a more significant fraction of the hot inner region is visible, leading to a more substantial X-ray component in the observed SED, especially if there is an optically thin hot corona above the disk. At super-Eddington accretion rates, although the total luminosity increases, the X-ray contribution diminishes because the hot regions are more obscured by the thicker disk (in Fig. \ref{fig:compilation}, notice that extreme values of the optical depth occupy a larger fraction of the solid angle in the X-ray weak case). This angle dependence implies that the apparent fraction of AGNs with intrinsic strong X-ray emission may be underestimated, as many may be viewed at angles where X-rays are suppressed. This is consistent with the explanation of the X-ray weakness provided by \cite{Maiolino_2024_Xray}, which requires significant covering factors. 
Additionally, the third row of panels in Fig. \ref{fig:compilation} shows the scattering optical depth out to $z = 100 r_g$, illustrating the transverse optical depth (integrated over $\theta$ from the pole at a constant radius). These results indicate that X-rays cannot escape laterally, suggesting significant opacity in the transverse direction of the jet.

The main differences in our super-Eddington simulations are shown in the temperature field (top row) and the $\beta \gamma$ field (bottom row), where $\beta = v/c$ and $\gamma$ is the Lorentz factor. The X-ray strong accretion structures display significantly higher values for the temperature and $\beta \gamma$, indicating that the jet is relativistic and more powerful. The combination of these two effects leads to enhanced X-ray emission.
Higher spin values generally lead to more efficient jet production, as the spin influences the inner accretion disk's structure and energy extraction efficiency (see \citealt{Ricarte_2023} for detailed results on jet efficiencies as a function of black hole spin for super-Eddington accretion).

The temperature distribution within the accretion disk is particularly revealing: the hottest regions, with $T > 10^8 \, \rm K$, are located at the base of the jet and on the wall of the funnel and are the primary sources of X-ray emission. Hence, observers aligned with the jet would predominantly see high-energy X-rays. Conversely, observers at larger inclination angles are screened from the hottest gas by the geometrically thick disk and see cooler regions, resulting in spectra dominated by UV and optical emissions.

In our GRRMHD simulations, the X-rays come from the base of the jet and the wall of the funnel, which is consistent with the lamppost interpretation of the X-ray ``coronae" in AGN (see, e.g., \citealt{Matt_1991, Martocchia_1996, Miniutti_2004}). Our simulations (similar to other studies using GRRMHD simulations) do not include an explicit treatment of the X-ray reflection component from an extended corona in the polar region.

\subsection{Explaining the X-ray Weakness of the LRDs}
\label{subsec:explaining_LRDs}

\begin{figure*}%
    \centering
\includegraphics[angle=0,width=0.90\textwidth]{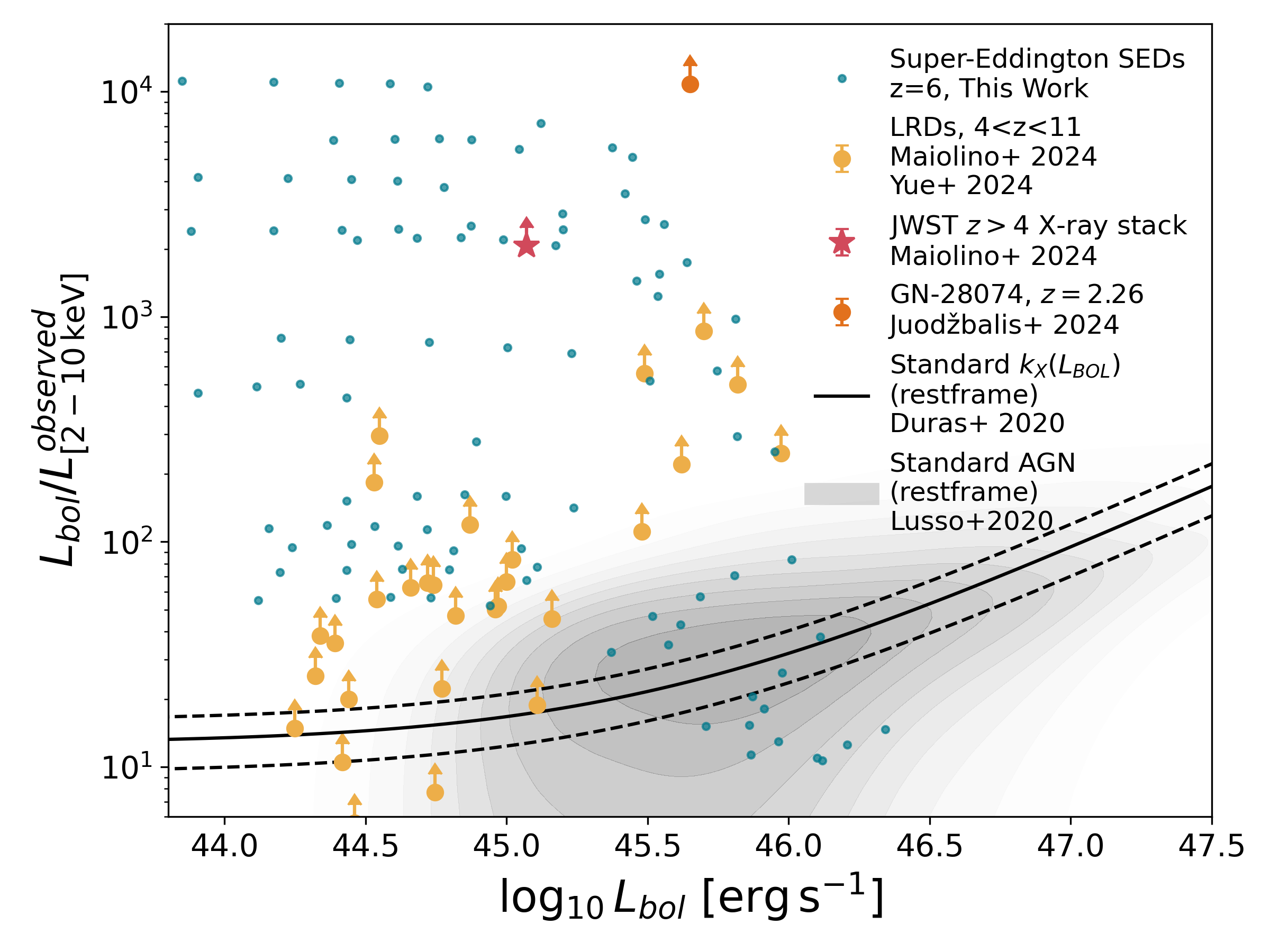} \hfill
    \caption{In the plane $\log_{10} L_{\mathrm{bol}}$ vs. X-ray bolometric correction (in the \textit{observed} $2-10$ keV range, for $z=6$), we compare the X-ray weakness of our super-Eddington SEDs with that of high-$z$ LRD populations, and some well-known low-$z$ samples.
    Teal dots represent the bolometric corrections derived in this study from super-Eddington SEDs at $z = 6$ (i.e., the median redshift from LRDs catalogs). Note that the bolometric luminosities of our SEDs are compatible with those observed. Orange circles indicate LRD populations at $4<z< 11$ from \cite{Maiolino_2024_Xray} and \cite{Yue_2024_Xray}, with arrows showing lower limits, while the red star marks the stacked data point (for sources at $z > 4$) from \cite{Maiolino_2024_Xray}, which provides the most robust constraint. The dark orange circle displays the location of GN-28074, which is the most X-ray weak among all AGN discovered by JWST ($z = 2.26$, \citealt{Juodzbalis_2024_extreme}). Note that each data point has been shifted to the \textit{observed} $2-10$ keV range to be comparable to our SEDs.
    The solid black line (with its spread) represents the \textit{restframe} $k_X(L_{\mathrm{bol}})$ relation from \cite{Duras_2020}. The grey distribution in the background shows low-$z$ Type 1 AGN data from \cite{Lusso_2020}.}
    \label{fig:LRD_Xray_BC}%
\end{figure*}

We can now apply our super-Eddington SEDs to the case of the LRDs.
Figure \ref{fig:LRD_Xray_BC} shows the bolometric correction $k_X^{z=6}$ in the \textit{observed} $2-10$ keV frame for our sample of super-Eddington SEDs, calculated at $z=6$. Note that the bolometric luminosities of our SEDs are compatible with those observed. We display the entire sample of $12$ simulations $\times$ $8$ viewing angles, totaling $96$ values of the bolometric correction.
These are compared to (i) lower limits on the bolometric corrections for LRDs at $4<z<11$ derived from \cite{Maiolino_2024_Xray} and \cite{Yue_2024_Xray}; (ii) the X-ray stacking of the \cite{Maiolino_2024_Xray} sample, which provides the most robust constraint; (iii) the standard (restframe) relation between the bolometric luminosity and the $2-10$ keV bolometric correction presented in \cite{Duras_2020}; (iv) a density distribution summarizing the (restframe) location in the $L_{\mathrm{bol}}-k_X$ plane of standard, low-$z$ Type 1 AGN from \cite{Lusso_2020}.
Note that the data points from \cite{Maiolino_2024_Xray} and \cite{Yue_2024_Xray} have been shifted to the \textit{observed} $2-10$ keV range to be comparable to our SEDs, using the redshift of each source, and $\Gamma=1.7$ and $\Gamma = 1.8$, respectively, as indicated in their study.

Also note that the LRDs are, in principle, a subset of the collection of Type 1 AGN at $4<z<11$ presented in \cite{Maiolino_2024_Xray}. The X-ray weakness is not exclusively associated with the LRDs but with a broader collection of JWST-discovered sources \citep{Maiolino_2024_Xray}. However, most sources in \cite{Maiolino_2024_Xray} are also classified as LRDs \citep{Maiolino_2023_new, Matthee_2023}. In addition, some of the LRDs presented in \cite{Yue_2024_Xray} are also included in \cite{Maiolino_2024_Xray}; for those not included, we calculated the bolometric luminosity from the provided $\Ha$ broad component luminosity, following the standard relation in \cite{Richards_2006}.

Figure \ref{fig:LRD_Xray_BC} demonstrates the following:
\begin{itemize}
    \item Many SEDs (i.e., $\sim 40\%$) are characterized by X-ray bolometric corrections higher than the lower limits on the single LRDs.
    \item Many SEDs (i.e., $\sim 35\%$) are characterized by X-ray bolometric corrections higher than even the lower limit on the stacked sample.
    \item Most SEDs (i.e., $\sim 88\%$) are characterized by values of the (observed) X-ray bolometric correction significantly higher than those of low-$z$ Type 1 AGN. Some SEDs have values of $k_X^{z=6}$ that are $2$ orders of magnitude higher than the highest (restframe) bolometric corrections measured by \cite{Duras_2020}, despite being associated with significantly (bolometrically) fainter objects.
\end{itemize}

The single SEDs shown in Fig. \ref{fig:LRD_Xray_BC} are not equally probable. In Sec. \ref{subsec:X_weakness}, we showed how most X-ray emission is associated with SEDs seen from a small (i.e., $< 30^\circ$) angle from the pole. Assuming random orientations (and two jets), the probability of observing such geometrical configuration is $\sim 13\%$. Hence, it is most likely to observe such systems along a line of sight that leads to an X-ray weak SED.

It is then apparent that the X-ray weakness observed in the LRDs can be explained by SEDs for mildly super-Eddington accreting SMBHs, typically characterized by lower or zero spin and observed away from the pole, which is the most likely observing condition.

Remarkably, our models predict that the restframe optical and UV spectra would still exhibit BLRs, potentially with higher velocities and implied line widths due to strong winds from the super-Eddington accreting SMBH. 
There is recent observational evidence of strong winds launched from super-Eddington accreting AGN \citep{Vietri_2020, Vietri_2022, Ding_2022_wind}, supported by robust investigations on the theoretical side \citep{Yang_2014, Sadowski_2015, Sadowski_2016, Yang_2023, Zhang_2024}. This implies that BLRs could appear artificially broadened by these winds, leading to an overestimation of the black hole mass and an underestimation of the Eddington ratio.

Of course, explaining the X-ray weakness of one specific source requires a detailed study involving GRRMHD simulations tailored for the particular (estimated) black hole mass and Eddington ratio. A study of a compilation of such SEDs, as a function of the spin parameter and viewing angle, will then associate a precise probability for that specific source to be X-ray weak.
However, our analysis, based on the median SMBH mass estimated ($\Mblack = 10^7 \Msun$) and the median redshift ($z\sim 6$), has demonstrated that this is a valid explanation for the X-ray weakness of the LRDs.
Thanks to the thick accretion disk produced by the super-Eddington accretion regime, our solution supports the requirement of large covering factors proposed by \cite{Maiolino_2024_Xray}.

Two final questions remain: (i) Are the LRDs possibly accreting at super-Eddington rates? (ii) Are the super-Eddington rates and the low spin required compatible?

\subsection{Eddington Ratios of the LRDs}
LRDs have a median Eddington ratio of $\fedd = 0.4$, \citep{Harikane_2023, Maiolino_2023_new, Larson_2023, Kokorev_2023}. Some LRDs are already estimated to be accreting at super-Eddington rates \citep{Maiolino_2023_new, Maiolino_2024_Xray, Harikane_2023}, while some Type 1 AGN seem to be well below the limit \citep{Ignas_2024}. Remarkably, some collections of LRDs, such as the one presented in \cite{Harikane_2023}, have a median Eddington ratio already super-Eddington: $\fedd = 1.1$.

As the bolometric luminosity is measured (or, better, inferred from bolometric corrections, such as the ones in \citealt{Richards_2006}), a decrease in the SMBH mass would automatically increase the Eddington ratio. 
Hence, assuming that the SMBH masses estimated for the LRDs are overestimated by a mere factor of $\sim 3$ would bring the median value of the Eddington ratio well inside the mildly super-Eddington regime, characterized by extreme X-ray weakness.
We note that a factor of $\sim 3$ is well within the uncertainty in the SMBH mass estimates via single-epoch virial relations even in the local Universe \citep{Kaspi_2000, Greene_2005}.

Hence, if the SMBH masses are systematically overestimated by a mere factor of $\sim 3$, our analysis for the median mass $\sim 10^7 \Msun$ at the median redshift of $6$ can explain the median population, i.e., $50\%$ of the sources.

\subsection{Spin-down Effect for Super-Eddington SMBHs}
\label{subsec:spin_down}
We have shown that mildly super-Eddington ($1.4<\fedd <4$) accreting SMBHs when slowly spinning ($a \sim 0$) and viewed far from the pole are extremely X-ray weak.

The combination of super-Eddington accretion rates and low spin is self-consistent: super-Eddington rates naturally lead to a spin-down of the black hole (see, e.g., recent studies by \citealt{Curd_2019, Ricarte_2023, Lowell_2024, Jacquemin-Ide_2024}). In particular, these studies have shown that the spin evolution and jet efficiency of an accreting SMBH are linked to the accretion rate.

A black hole accreting from a disk with a strong poloidal magnetic field (i.e., a MAD accretion disk) is characterized by a jet, which is powered by the spin energy of the black hole. 
The spin-down effect becomes crucial as the accretion rate surpasses the Eddington limit. Detailed GRRMHD simulations indicate a dramatic reduction in the black hole's spin in the super-Eddington regime, driving it toward a near-zero spin state \citep{Ricarte_2023}.
In particular, the spin-down process, for the range of Eddington ratios investigated here (i.e., $1.4 < \fedd < 13.4$), occurs on cosmologically-negligible timescales of $3-30$ Myr \citep{Ricarte_2023}.

Hence, if the SMBHs hosted by the LRDs are accreting at a super-Eddington rate due to the large availability of cold gas \citep{Pacucci_2020}, their spin is expected to reach near-zero equilibrium levels quickly, further enhancing their X-ray weakness.
Thus, with the LRDs, we may witness, for the first time, a ubiquitous occurrence of slowly spinning SMBHs accreting at mildly super-Eddington rates.

\section{Summary and Conclusions}
\label{sec:conclusions}
This study was motivated by the widespread JWST detection at $z > 4$ of sources characterized by solid evidence for the presence of low-luminosity AGN, with SMBH masses of $\Mblack=10^{6-8} \Msun$. These SMBHs, found in compact and red hosts (the LRDs), are largely undetected in X-rays, leading to the ``X-ray weakness'' problem.

Using a suite of GRRMHD simulations, we investigated the super-Eddington accretion process onto a SMBH with mass $\Mblack = 10^7 \Msun$ at $z \sim 6$; these values represent the medians for the LRDs population in the largest catalogs thus far available \citep{Harikane_2023, Maiolino_2023_new, Kocevski_2024, Akins_2024, Kokorev_2024_census}.
Our key findings are summarized here.
\begin{itemize}
    \item The highest levels of X-ray weakness occur in SMBHs accreting at mildly super Eddington rates ($1.4<\fedd<4$), with low or zero spin ($a\sim 0)$, viewed at angles $>30^\circ$ from the pole. Viewing angles aligned with the pole increase the observed X-ray emission.
    \item X-ray bolometric corrections in the \textit{observed} $2-10$ keV energy band reach values of $\sim 10^4$ at $z \sim 6$, which is $\sim 5$ times higher than the highest constraint from X-ray stacking. These X-ray corrections are compatible even with the most extreme X-ray weak sources detected. About $35\%$ of our SEDs have X-ray bolometric corrections higher than limits imposed by X-ray stacking.
    \item About $55\%$ of our super-Eddington SEDs have optical-UV to X-ray ratios ($\alphaox$) outside the typical range for standard Type 1 AGN. Contrary to the standard, our SEDs show increasing X-ray emission for increasing optical-UV luminosity.
    \item Our super-Eddington SEDs are extraordinarily steep and soft in the X-rays. Their median photon index is $\Gamma = 3.1$, while the most common value is $\Gamma=4.4$. In $86\%$ of our SEDs, $\nu F_\nu$ declines rapidly with increasing frequency. These values resemble the highest measured in NLSy1 galaxies. 
    \item X-ray strong SEDs have a harder X-ray spectrum with a bump at $10^{19} - 10^{20}$ Hz (restframe) when viewed close to the pole (i.e., $<30^\circ$), while X-ray weak SEDs lack this bump except, minimally, for a viewing angle of $\sim 10^\circ$.    
    \item If X-ray weak spectra characterize the population of LRDs, Chandra may detect only those viewed from minimal angles from the jet. The probability of observing a SMBH within $10^\circ$ from the jets is $\sim 1.5\%$, which is similar to the X-ray detection rate ($\sim 0.6\%$) in the most extensive catalog of LRDs with X-ray coverage available \citep{Kocevski_2024}.
    \item Next-generation X-ray observatories, such as AXIS, can detect typical X-ray weak sources at $z\sim 6$ from any viewing angle.
    \item The effect of the viewing angle, or beaming, is crucial. The same SMBH, with the same Eddington ratio and spin, produces SEDs whose peaks differ by $1.5$ dex in flux when the viewing angle changes from $80^\circ$ to $10^\circ$. In general, 
    observers aligned with the jet axis are most likely to see X-rays. Observers at larger and more probable viewing angles see cooler regions, resulting in spectra dominated by optical-UV emission.
    \item The inner structure of the accretion flow reveals the physical difference between X-ray weak and strong SEDs. Enhanced X-ray emission is linked to higher temperatures ($T > 10^8$ K) close to the SMBH and a highly relativistic jet, with higher spin values leading to more powerful jets and harder X-ray spectra.
    \item In the super-Eddington regime, while the total luminosity increases, the X-ray contribution diminishes because the hot regions are more obscured by the thicker disk. This picture is consistent with the model by \cite{Maiolino_2024_Xray}, which requires significant covering factors. Thus, our proposal consistently complements other models to explain the lack of X-rays.
    \item Super-Eddington accretion rates naturally lead to strong winds, artificially broadening BLRs and potentially overestimating the black hole masses.
    \item Our picture is fully self-consistent: super-Eddington rates lead naturally to lower spin, large covering factors, and broad emission lines, leading to an intrinsic X-ray weakness. The spin-down process occurs over $3-30$ Myr for the range of Eddington ratios investigated.
    \item Many SMBHs in the LRDs are already estimated to accrete at super-Eddington rates. If the SMBH masses are systematically overestimated by a mere factor of $\sim 3$ (which is compatible with even local measurements), our analysis for the median mass $\sim 10^7 \Msun$ at the median redshift of $6$ can explain the median population, i.e., $50\%$ of the sources.
    \item The phenomenon of the LRDs is mainly observed in the redshift range $4 < z < 8$, where large availability of accretable gas would render widespread super-Eddington accretion feasible. An older Universe, with far less free gas available, would make this phenomenon disappear, as observed.
\end{itemize}

A comprehensive understanding of the masses, accretion rates, and spins of the SMBHs in the LRDs is crucial for enlightening their role in the early formation and evolution of their hosts. Specifically, accurate measurements of the black hole and the host's stellar mass are essential. Recent analyses indicate that the central SMBHs in LRDs appear to be overmassive by a factor of $10-100$ relative to the stellar content of their hosts \citep{Pacucci_2023_JWST, Pacucci_2024_z_evolution, Durodola_2024}, according to standard local relations (e.g., \citealt{Reines_Volonteri_2015}). If confirmed, this discrepancy could provide the most substantial evidence yet for the occurrence of heavy seed formation in the high-$z$ Universe \citep{Agarwal_2013, Natarajan_2017, Scoggins_2023, Pacucci_2024_z_evolution}.
For this and other reasons, it is essential to constrain the mass and accretion properties of the SMBHs detected in the LRDs.

In summary, our study suggests that with the LRDs we are witnessing, for the first time, the ubiquitous occurrence of slowly spinning SMBHs accreting at mildly super-Eddington rates.

\vspace{10pt}
\noindent \textit{Acknowledgments:} F.P. acknowledges fruitful discussions with Nico Cappelluti, Andrea Ferrara, Erin Kara, Vasily Kokorev, Roberto Maiolino, Joseph Silk, and Minghao Yue.
F.P. also acknowledges support from a Clay Fellowship administered by the Smithsonian Astrophysical Observatory. This work was also supported by the Black Hole Initiative at Harvard University, funded by grants from the John Templeton Foundation and the Gordon and Betty Moore Foundation. 
This work used the Anvil supercomputer at Purdue University through allocation AST-080028 from the Advanced Cyberinfrastructure Coordination Ecosystem: Services \& Support (ACCESS) program, which is supported by National Science Foundation grants \#2138259, \#2138286, \#2138307, \#2137603, and \#2138296.

\bibliography{ms}
\bibliographystyle{aasjournal}



\end{document}